\documentclass[aps,twocolumn,prl,superscriptaddress,showpacs]{revtex4}
\usepackage{epsfig,amsmath,amsbsy,amssymb}
\usepackage{graphicx,color}

\begin{document}

\title{Slow Crack Propagation  in  Heterogeneous Materials}
\author{J. Kierfeld}
\affiliation{Max Planck Institute of  Colloids and Interfaces,
Science Park Golm, 14424 Potsdam, Germany}
\author{V.M. Vinokur}
\affiliation{Argonne National Laboratory,
Materials Science Division,
9700 South Cass Avenue,
Argonne, Illinois 60439, USA}
\date{\today}

\begin{abstract}
Statistics and thermally activated
dynamics of crack nucleation and  propagation in a two-dimen\-sional
heterogeneous material containing {\em quenched randomly
distributed} defects are studied theoretically. 
Using the generalized Griffith criterion
we derive the equation of motion for the crack
tip position accounting for dissipation, thermal noise and the
random forces arising from
the defects. We find that aggregations of defects
generating  long-range interaction forces (e.g.,  clouds of
dislocations) lead to anomalously slow creep of the crack tip or
even to its complete arrest.
We demonstrate that  heterogeneous materials with frozen defects
 contain a large number of arrested microcracks and that their
 fracture toughness
is enhanced to the experimentally accessible time scales.
\end{abstract}



\maketitle

\begin{figure}
\begin{center}
  \epsfig{file=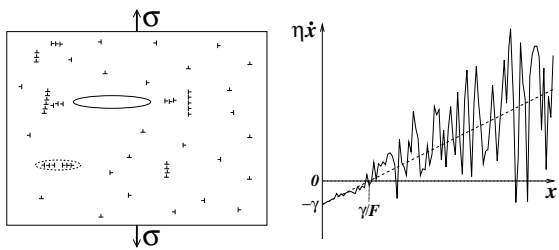,width=0.41\textwidth}
\caption{\label{crackseps}
Left: Sketch of an arrested crack 
(solid ellipse) in a random array of frozen dislocations (with cores
 represented by symbols ${\bf \perp}$);
the dashed ellipse indicates a favorable  region for crack nucleation.
 Right:
Typical realization of the
effective  force $-\gamma+Fx+f_d(x)$, see Eq.\ (\ref{EOM}),
 acting on the crack tip for short-range correlated random forces
$f_d(x)$ with $\delta=3$. }
\end{center}
\end{figure}

Fracture mechanisms and their relation to the material structure
is a long-standing problem \cite{lawn}. Ideal crystals are subject
to fast brittle fracture -- as was first explained by Griffith
\cite{G20} -- while homogeneously amorphous systems exhibit slow
ductile fracture controlled by the plastic deformation at the
crack tip \cite{lawn,freund}. Real materials are neither of the
above.  Real crystals do contain defects; but even in the
ultimately disordered substances the defects are not distributed
homogeneously but form spatially inhomogeneous aggregates such as
inclusion clusters and/or dislocation pileups
\cite{lawn,HR90,AN01}. It is intuitively plausible
that defect aggregates promote crack nucleation as the crack 
can settle at an
energetically favorable nucleation site. At the same time one can
expect that random heterogeneities impede the subsequent crack
propagation process as shown in  Fig.\ \ref{crackseps}. This poses
the important question about the ultimate 
 effect of frozen inhomogeneities on the fracture mechanism and, in
particular, whether frozen defects enhance or degrade the
fracture toughness of a material.

The inspiring work \cite{AN01} discussed the disorder-stimulated
{\em nucleation} of  critical cracks. In this Letter, we focus on
the {\em dynamics} of cracks in heterogeneous materials and
investigate fracture probabilities and the statistics of the
fracture times. We consider {\em both} zero- and 
finite-temperature crack dynamics, the latter being governed by thermal
activation. We restrict ourselves to the simplest case of cracks
in a thin (quasi-) two-dimensional (2D) ideally elastic plate
containing random heterogeneities. In two dimensions a crack front
is a point -- the crack tip; thus the additional effects arising
from  crack front roughening are absent. Building on 
Griffith's concept of  energy balance \cite{G20}, we consider
 crack tip motion governed by the dynamic energy release rate
\cite{freund,FM99} and derive the equation of motion for the crack
tip.  We include both dissipative and thermal forces, and the
position-dependent random forces acting on the crack tip due to
frozen material defects. We discuss three basic kinds of frozen
inhomogeneities: (i) bond strength variations, (ii) random
impurities resulting in  local compression of the elastic
medium, and (iii) frozen dislocations interacting with the crack.
The range of elastic interactions with the crack tip increases
when going from the type (i), to types (ii) and (iii).

We find that the supercritical cracks can be  {\em arrested} by
heterogeneities inducing  long-range elastic forces, i.e.,  by
 frozen dislocations (iii).  We show that thermally activated
cracks exhibit {\em anomalously slow dynamics} with  vanishing
mean velocity for all three types of disorder. We conclude that
quenched defects effectively slow down the crack propagation and
derive experimentally observable characteristic material
properties such as the statistics of the critical stresses and the
power-law distributions of crack waiting times.  
This explains 
that in materials containing long-range structural defects 
arrested microcracks are experimentally observable 
\cite{Ga97}.

\paragraph{Crack tip equation of motion.}

Let a single planar crack extend from $-x/2$ to $+x/2$ along the
$x$ direction of a 2D elastic medium of  size $L$ loaded in 
mode I by a uniform external stress $\sigma$. In  a perfectly
homogeneous elastic medium the Griffith crack energy is the sum of
the elastic energy gain $E_{el}(x)$ and the crack surface energy
$E_{s}(x)$. The driving force for the crack tip advance is the
release of the elastic energy quantified by the {\em static energy
release rate}.
$G(x) =  - \partial_{x} E_{el} =  \pi \sigma^2 x/2Y \equiv F x$ \cite{lawn},
where
$Y=E/(1-\nu^2)$ is the 2D Young's modulus.
The energy release is balanced by the {\em specific fracture  energy}
$\gamma$, related to the crack surface energy $E_s$ by 
 $\gamma= \partial_{x} E_{s}$.
Griffith's criterion for the onset of crack growth gives
   $G> G_c = \gamma$ \cite{G20}, where
$G_c$ is the {\em critical energy release rate} that can be reached by
increasing  the  crack length beyond the critical value $x_c =
\gamma/F = 2\gamma Y/\pi\sigma^2$.

Material heterogeneities are described by the frozen random forces
$f_d(x)$ which are included into the Griffith's force balance. We
adapt Gaussian distributed random forces with  zero mean value,
$\overline{f_d(x)}=0$, where the overbar denotes  the average over
disorder, and consider two types of forces:
Short-range
correlated forces  (SRCF)
$\overline{f_d(x)f_d(x')} = \Delta_\delta x^\delta \delta_a(x-x')$
(where $a$ is a  microscopic cutoff length)
and long-range correlated forces  (LRCF)
$\overline{f_d(x)f_d(x')} = \Delta_\delta x^{(\delta-1)/2}
{x'}^{(\delta-1)/2}$.
$\Delta_\delta$ is
 the strength of the random forces proportional
to the defect concentration,
 and the exponent $\delta$ characterizes the
elastic interaction between frozen defects and the crack. Note
that for both types of random forces the corresponding
potential energies $E_d(x)$ of the crack tip defined by the
relation $f_d(x) = -\partial_x E_d$ 
have  the same correlations
$\overline{(E_d(x+z) - E_d(x))^2} \sim \Delta_\delta z^{\delta+1}$
for $z \gg x$. It was shown in Ref.~\cite{AN01} that random bonds
(random fracture toughness) result in SRCF  with $\delta=0$,
impurities produce SRCF with $\delta=1$, and dislocations induce
LRCF with $\delta=3$, i.e., $\delta$ increases with the range of
the the elastic interaction between the crack tip and defects
\cite{remark1}.

Dissipation occurs mostly near the crack tip, where the elastic
energy transforms into heat via  plastic deformation
\cite{FM99}, and can thus be described as the local viscous force
exerted on the tip, $-\eta \dot{x}$,
 with $\eta$ being the tip  viscosity.
Including the thermal force $\zeta(t)$ 
acting on the crack tip into the force balance,
we obtain the overdamped equation for the crack tip motion as
 $\eta \dot{x} = G(x) -\gamma  +f_d(x) +\zeta(t)$.
As we focus on slow crack dynamics thermal fluctuations
 facilitate both transient crack healing and extension, and the system is 
 close to thermal equilibrium  such that it is justified to use $\langle
\zeta(t) \rangle =0$ and 
correlations 
 $\langle \zeta(t) \zeta(t') \rangle = 2\eta T \delta(t-t')$ 
($k_B \equiv 1$) for thermal forces. 
In order to complete the description of dynamics,
  the kinetic energy
 of the elastic medium has to be
taken into account via the
 {\em dynamic energy release rate} $G(x,\dot{x})$ \cite{freund}:
 \begin{equation}
 G(x,\dot{x}) = A(\dot{x}) G(x)  \approx
    \left(1-\dot{x}/c_R\right)  G(x)
~.
\label{Gv}
\end{equation}
In general, $A(\dot{x})$ decreases  monotonically  with
increasing crack tip velocity $\dot{x}$, starting with  $A(0)=1$
in the static limit and reaching zero $A(c_R)=0$ at
the  Rayleigh wave velocity $c_R$
 \cite{freund}. The approximation in Eq.\ (\ref{Gv})
complies with most experiments and will be used in what follows.
In a homogeneous material the dynamic force balance $G(x,\dot{x})
= \gamma$ generalizes the Griffith criterion and describes the
energy flux into the crack tip and its subsequent conversion into
crack surface energy. Including the viscous force $-\eta
\dot{x}$, the thermal force  $\zeta(t)$, and the
 frozen random forces $f_d(x)$
 into  the dynamic force balance,
$-\eta \dot{x} +G(x,\dot{x}) = \gamma  - f_d(x) - \zeta(t)$, one
finally arrives at the following equation of motion:
\begin{subequations}
\begin{eqnarray}
  \eta \dot{x}
  &=&  B(x)[-\gamma +Fx +f_d(x) +\zeta(t) ]
 \label{EOM}\\
  B(x) &\equiv &[1+ (\gamma/\eta c_R)(x/x_c) ]^{-1}
 \equiv  [1+ x/b ]^{-1},
\label{B}
\end{eqnarray}
\end{subequations}
which is a Langevin equation with multiplicative noise
\cite{kampen};
$b \equiv  \eta c_R/F$ is the characteristic crack length.
 For small cracks $x \ll b$, $B(x) \approx 1$, and Eq.\
(\ref{EOM}) reduces to the usual  overdamped dynamics. On the
other hand, for large cracks, $x \gg b$, $B(x)$ vanishes as $B(x)
\approx b/x$ giving rise to a low effective temperature.

\paragraph{Crack nucleation and propagation at $T=0$.}

In the absence of flaws [$f_d(x)=0$], cracks can  be thermally
nucleated at temperatures  $T>0$ \cite{thnucl}, but the material
is stable against  crack formation at $T=0$. However,  there is
a non-zero probability that cracks of the critical size $x \sim
x_c$ are ``nucleated''  by  quenched disorder even at $T=0$
 if the typical
disorder energy gain $\overline{E_d^2(x_c)} \sim \Delta_\delta
 x_c^{1+\delta}$, which scales in the same way for both SRCF and LRCF,
compensates for the nucleation energy  $\Delta E_c = \gamma^2/2F$
 \cite{AN01}.
The resulting probability for  disorder-induced nucleation,
\begin{equation}
p_{\rm nucl} = {\rm prob}[\Delta E_c\!+\! E_d(x_c)\!<\!0]
  \sim  e^{
   -\gamma^{3-\delta}F^{\delta-1}/ \Delta_\delta },
\label{pnucl}
\end{equation}
increases with the increasing disorder strength
$\Delta_\delta$.

After its nucleation,  a growing crack can get arrested by frozen
disorder which, thus, can prevent fracture. Balancing the driving
force and the typical random force developing over the distance
$x$, $(\overline{f_d(x)^2})^{1/2} \sim
(\Delta_\delta/a)^{1/2}x^{\delta/2}$ for SRCF, we find a crack
{\it arresting length} $x^* \equiv
(F^2a/\Delta_\delta)^{1/(\delta-2)}$ which characterizes the {\em
static} force equilibrium (we focus on
 supercritical cracks $x \gg x_c$ and neglect the  crack surface
 energy $\gamma$). For $\delta <2$,
the force equilibrium is unstable against  further propagation
giving rise to brittle fracture,
whereas it is stable for $\delta >2$ because
sufficient  driving forces become exponentially rare at
$x>x^*$, which leads to a ductile fracture mechanism. 
For LRCF,  we find an arresting length $x^* \equiv
(F^2/\Delta_\delta)^{1/(\delta-3)}$ and the force equilibrium becomes
stable for  $\delta >3$.

The probability $p_{\rm prop}$ for the crack propagation at $T=0$
is given by the probability of finding a positive force on the
crack tip for {\em all} $x>x_c$, $p_{\rm prop} \approx
 \prod_x  {\rm prob}[Fx\!+\!f_d(x)\!>\!0]$ \cite{AN01}.
SRCF show  two qualitatively different behaviors depending on
the value of $\delta$:
\begin{equation}
 p_{\rm prop} \sim
 \left\{
  \begin{array}{ll}
   e^{ - x^*/a} \sim e^{- (F_a/F)^{2/(2-\delta)}}, & \delta <2
  \\
        \left[ 1-e^{-(F/F_L)^2} \right]^{L/a}, 
       & \delta \ge 2 
  \end{array}
\right.
\label{pprop2}
\end{equation}
where $F_a \equiv (\Delta_\delta a^{\delta-3})^{1/2}$ and $F_L
\equiv (\Delta_\delta L^{\delta-2}/a)^{1/2}$ are the
characteristic forces, defined by  $x^*=a$ and  $x^*=L$,
respectively. For $\delta<2$ the fracture mechanism is
brittle, as  $p_{\rm prop}$ is independent of the system size $L$ 
and strongly increases if  $x^*$ becomes
comparable to the microscopic length $a$, whereas for $\delta\ge
2$, the fracture is ductile because $p_{\rm prop}$ decreases for large
$L$ and becomes appreciable 
only if the stable crack of length $x^*$ becomes comparable to $L$. 
For LRCF, on the other hand, we find
\begin{equation}
p_{\rm prop} \sim
  1- e^{- \left( F/\tilde{F}_L\right)^2 }
\label{pprop2L}
\end{equation}
with  $\tilde{F}_L \equiv \left( \Delta_\delta L^{\delta-3}
\right)^{1/2}$, which lacks an exponential dependence on $L/a$
because the arresting forces are strongly correlated.
The case of  frozen dislocations (iii) with $\delta=3$
is marginal.
 Cracks are arrested for disorders $\Delta_3 \gg
 F^2$ or up to high strain levels
$\sigma/Y \ll  b_{\rm (fd)}c_{\rm (fd)}^{-1/2}
   \sim 0.05$ where $b_{\rm (fd)}$ and $c_{\rm (fd)}$ are Burgers
   vector and 2D concentration of dislocations and we used an estimate
for strong disorder in glass from Ref.\ \cite{AN01}. Fracture will
occur if a crack both nucleates and propagates with the
probability $p_{\rm frac} = p_{\rm nucl}p_{\rm prop}$. For SRCF
with $\delta<2$, i.e., random bonds (i) or random impurities (ii),
the fracture probability is {\em finite} for large $L$. For LRCF
with $\delta=3$, i.e., frozen dislocations (iii), we find a {\em
complementary} behavior $p_{\rm nucl} \sim 1- p_{\rm prop}$, which
enhances fracture toughness because at small defect densities the
nucleation is unlikely, whereas at large defect concentrations the
propagation is suppressed.

The fracture probability (as a function of $F$) equals the
probability that $F$ is larger than the critical  force $F_c\equiv
\max_x\{[\gamma-f_{d}(x)]/x\}$ of the given crack, i.e., ${\rm
prob}[F_c\!<\!F]=p_{\rm frac}(F)$. For SRCF, cracks can nucleate
at $N\sim (L/x_c)^2$ statistically independent seed locations in
the sample \cite{AN01} with critical forces drawn from the
distribution $p_c(F_c) = \partial_F p_{\rm frac}(F_c)$. Then the
fracture occurs at the ``weakest link'' if the applied load $F$
exceeds the {\em smallest} critical force of all $N$ crack nuclei.
The  {\em average}
 fracture force $\overline{F_{\rm frac}}$  follows from the
condition $1=N p_{\rm frac}(\overline{F_{\rm frac}})$. For
$\delta<2$,  we find
$\overline{F_{\rm frac}} \sim
F_a(\ln N)^{-(2-\delta)/2}$.
Because  $p_{\rm frac}(F_c)$ and thus  $p_c(F_c)$
decrease exponentially with $1/F_c$,
see eq.\ (\ref{pprop2}),
the resulting {\em distribution} of  fracture forces for
$\delta<2$ is an extreme value distribution of the Gumbel type,
\begin{equation}
  {\rm prob}[F_{\rm frac}\!>\!F] \sim \exp[-c_1 N e^{-c_2
  (\ln N)^{\delta/2} F_a/F}   ]
\label{gumbel}
\end{equation}
with constants $c_1$ and $c_2$. 
This result applies to random bonds (i) and random impurities (ii)
and  generalizes previous findings
 for random fuse models \cite{rf_extreme},
which correspond to the special case of  random bonds (i) with $\delta=0$.
The probability
${\rm prob}[F_{\rm frac}\!>\!F]$  in (\ref{gumbel}) equals
 the probability that the sample will not fracture
and, thus, that all $N$ statistically independent cracks are
arrested.
For LRCF, on the other hand, crack energies at different positions
also have  long-range correlations; thus extreme value statistics
of fracture probabilities does not emerge and the fracture
probability is simply given by $p_{\rm frac} = p_{\rm nucl}p_{\rm
prop}$.

\paragraph{Dynamics of  thermally activated crack propagation.}

Having established the conditions and probabilities for the crack
arrest by heterogeneities at $T=0$,  we address the question of to
what extent these findings have to be modified by thermal
fluctuations. While at $T=0$, any energy barrier leads to crack
arrest, thermal fluctuations at $T>0$ give rise to activated crack
propagation. The equation of motion (\ref{EOM}) for the crack tip
resembles those for the overdamped motion of a  particle driven
over the one-dimensional disorder potential $E_d(x)$ extensively
reviewed in Ref.\ \cite{BG90}. At low temperatures, the particle
exhibits slow dynamics due to the wide distribution of energy
barriers giving rise to anomalously slow diffusion, creep, or even
particle arrest \cite{DV95}.

We consider an ensemble of cracks arrested at $T=0$ by the random
forces and address the question whether it stays arrested when the
finite temperature, $T>0$, is switched on. To this end we analyze
the dynamics of a {\em typical} crack (since it is the typical
crack that is arrested at $T=0$, and propagating cracks represent
{\em rare} events as follows from the previous section), which is
described by the Fokker-Planck equation for the probability
density $P(x,t)$ corresponding to the equation of motion
(\ref{EOM}) \cite{ito}:
\begin{eqnarray}
 \partial_t P &=& -\partial_x J\\
  \eta J
  &=&  -T \partial_x[B^2(x)P] + B(x)[ -\gamma +Fx+f_d(x)] P~~~~
  \label{J}
\end{eqnarray}
After
finding stationary solutions $P(x)$ for a  non-zero constant
current $J$ and absorbing boundary conditions $P(L)=0$, the
normalization condition $\int_0^L dxP(x)=1$ determines the fracture
time $\tau_{\rm frac} = 1/J$ \cite{MFP},
\begin{eqnarray}
 \overline{\tau_{\rm frac}} &\approx &
    \int_0^L \!dx \frac{\eta}{T B^2(x)} \int_0^{\infty}\! dz
    e^{- I(x,z)}~~~\mbox{with}
  \nonumber\\
 I(x,z)  &\equiv &
   \int_x^{x+z} du \left[ Fu/TB(u) - \Delta_\delta
   u^{\delta}/T^2B^2(u)  \right]~~~
\label{taufr}
\end{eqnarray}
where we  took the limit of infinite $L$ and averaged over
disorder. The behavior of $I(x,z)$ for large $z$
governs the fracture time and is identical for both SRCF and LRCF
\cite{ito}. Using the asymptotics for large $u$,
$B(u) \approx  b/u$, see Eq.\ (\ref{B}),  we find
a finite  mean fracture time
$\overline{\tau_{\rm  frac}}$  for $\delta<0$,
 whereas it diverges for $\delta >0$.
For $\delta=0$, the mean fracture time is infinite for $\Delta_0>
\Delta_{0,c}\equiv  FTb=T\eta c_R$, i.e.,  above the threshold
disorder strength $\Delta_{0,c}$, which is independent of the
driving force $F$.

Now we derive the distribution of random energy barriers that
govern the activated dynamics \cite{FV88}. As follows from the
equation of motion (\ref{EOM}), the effective random energy
controlling thermal activation  is $\phi(x)$ with $\partial_x\phi
\equiv (-Fx-f_d(x))/B(x)$. Therefore, we have to find  the
distribution of barriers $p(E)$ developing in the random energy
landscape $\phi(x)$ of a particle located initially at $x=x_i$.
This distribution evolves from the Gaussian distribution of random
forces $f_d(x)$ and can be written as  path integral in the
``energy space'' applying the formalism that has been developed in
Ref.\ \cite{FV88}. After some algebra, we finally find the
effective barrier distribution
\begin{equation}
 p(E) \sim e^{-{\rm const} (E/E_0)^{1-\delta/3}}
\label{pE}
\end{equation}
for large cracks  ($x\gg x_i$ and $x\gg b$) both for SRCF and
LRCF, where $E_0 = b^{-1}
\Delta_\delta^{3/(3-\delta)}F^{-(3+\delta)/(3-\delta)}$ is the
characteristic barrier energy.  The distribution (\ref{pE})
attains a simple exponential form for $\delta=0$. For $\delta>3$
large barriers are no longer rare, and $p(E)$ cannot be
normalized. Consider the crack tip starting at $x_i$ and traveling
over a distance $x_t \gg x_i$ for the time $t$. For $0\le
\delta<3$, the dynamics is controlled by the highest barrier,
$E_t$, it meets, obtained from the condition $1= (x_t/a)
\int_{E_t}^{\infty} dE p(E)$. Then $E_t \sim E_0
\ln^{3/(3-\delta)} x_t$, and  using the Arrhenius relation $t \sim
e^{E_t/T}$ we find:
\begin{equation}
  x_t \sim \exp[(T/E_0)^{1-\delta/3} \ln^{1-\delta/3} t ]
~~~\mbox{for}~0\le \delta<3.
\label{xt}
\end{equation}
For $\delta=0$, i.e., random bonds (i), this
represents anomalously slow diffusion with the power-law
 dynamics $x_t\sim t^{T/E_0}$ where  $E_0=\Delta_0/Fb$.
For $T<E_0$ or above  the threshold disorder strength
$\Delta_{0,c}$, we find the vanishing mean velocity $x_t/t \approx
0$ in agreement with our above result of a diverging  mean
fracture time $\overline{\tau_{\rm  frac}}$ for $\delta=0$;
$\delta<0$ leads to fast brittle fracture $x_t \sim t$ agreeing
with our conclusion about the corresponding {\em finite} mean
fracture time. For $\delta >0$ the crack motion law is slower than
any power with $x_t/t \approx 0$, representing the effective crack
arrest (meaning an infinite mean fracture time). For $\delta\ge
3$, the complete crack arrest, $x_t \approx 0$, occurs. Thus an
ensemble of cracks that was arrested at $T=0$ remains effectively
arrested (in the sense of an  infinite fracture time or zero
average velocity) for heterogeneities with $\delta>0$,  which
include frozen dislocations (iii) and random impurities (ii). For
random bonds (i) with $\delta=0$ we find the anomalously slow
diffusion with the power-law dynamics.

\paragraph{Conclusion and discussion.}

We have derived the equation of motion (\ref{EOM}) for the crack
tip by incorporating effects
from  dissipation, thermal fluctuations, and frozen
heterogeneities into the dynamic fracture criterion  $G(x,\dot{x})
= \gamma$. The tip equation of motion is an overdamped
Langevin-type equation for a particle in a one-dimensional
disordered potential. We have obtained the conditions and
probabilities for  crack arrest at zero temperature as a
function of the applied stress $F$ and the type of heterogeneities
involved as described by the exponent $\delta$ \cite{remark1} both
for the short-range (SRCF) and long-range (LRCF) correlated
forces. For SRCF, we find  complete crack arrest for $\delta
\ge 2$ and the extreme value probability (\ref{gumbel}) of 
crack arrest for $\delta<2$. For LRCF, cracks get arrested for
$\delta>3$,  and the  fracture probability is drastically reduced
for hard-worked materials \cite{workhardening} containing frozen
dislocations (iii) ($\delta=3$). Cracks that are arrested at $T=0$
can propagate by thermal activation at finite temperatures $T>0$.
For heterogeneities with $\delta\ge 0$, i.e., also for random
fracture toughness  (i) and impurities (ii), the disorder
potential leads to the slow crack dynamics (\ref{xt}) with 
zero mean velocity as the crack tip gets trapped in the deep
potential minima. This trapping mechanism is much more efficient
than the crack capture by crystal lattice effects \cite{sch90}
and explains the existence of  arrested metastable
 {\em microcracks} in heterogeneous materials with sizes that
 can be  considerably larger than the critical crack length of
the homogeneous material; this effect has been observed in a
number of recent experiments \cite{Ga97}. Experimentally observed
fracture precursors in heterogeneous materials with  power-law
waiting time distributions \cite{Ga97,G02,S02} can also be
explained  in the framework of our theory as characteristics of
the case $\delta=0$ of random fracture toughness (i).
It remains an open question for future investigations whether the
ensembles of arrested microcracks  become unstable with
respect to microcrack coalescence and  slow crack growth by
cyclic loading in fatigue experiments.

\paragraph{Acknowledgments.}

This research is supported by the US DOE Office of Science under contract
No. W-31-109-ENG-38.



\begin{thebibliography}{99}
\bibitem{ito}
We use
the Ito formulation \cite{kampen}, results  for   the
Stratonovich formulation are similar as
waiting times are governed by the same exponent
$I(x,z)$ in (\ref{taufr}).
\end{thebibliography}
\end{document}